# Establishing the Minimal Clinically Important Difference (MCID) for Smartphone-Derived Gait Measures in Multiple Sclerosis

Mike D Rinderknecht,[a] Bernhard Fehlmann,[a] Dimitar Stanev,[a] Cedric Simillion,[a] Ernst Bos,[a] Letizia Leocani,[b,c] Agne Kazlauskaite,[a] Gary Cutter,[d] Helmut Butzkueven,[e,f] Licinio Craveiro[a]

[a] F. Hoffmann-La Roche Ltd., Basel, Switzerland

[b] Faculty of Medicine, University Vita-Salute San Raffaele, Milan, Italy

[c] Scientific Institute IRCCS San Raffaele, Milan, Italy

[d] Department of Biostatistics, University of Alabama at Birmingham, Birmingham, AL, USA

[e] Department of Neuroscience, School for Translational Medicine, Monash University, Melbourne, Australia

[f] Department of Neurology, Alfred Health, Melbourne, Australia

**Corresponding author**

Mike D Rinderknecht

mike.rinderknecht@external.roche.com

# Abstract

**Background:** Digital health technologies allow for frequent, remote gait monitoring in people with multiple sclerosis (MS). However, to differentiate daily variability from actual disease progression in longitudinal data, established minimal clinically important differences (MCID) are required. Currently, there is limited literature defining these thresholds for digital gait metrics.

**Objective:** To establish MCIDs for digital gait measures reflecting progression in MS.

**Methods:** Digital gait measures were captured via daily, remote, smartphone-based Two-Minute Walk Tests in CONSONANCE (NCT03523858), a phase 3b study of ocrelizumab in progressive MS. Using an anchor-based approach, median changes from baseline at Week 96 on digital gait measures were computed for patients showing clinically meaningful worsening on either Timed 25-Foot Walk, Ambulation Score, Expanded Disability Status Scale, or 12-item Multiple Sclerosis Walking Scale. These changes were subsequently triangulated to derive the MCID estimates.

**Results:** 243 patients with progressive MS (female: n=125 (51%); mean [SD] age: 49.3 [9.3]; mean [SD] EDSS: 4.8 [1.4]) had digital gait data available at baseline and Week 96. Median changes were generally consistent across anchors. Triangulated MCIDs are: Step Velocity = -0.16 m/s, Step Velocity Scaled to Walking Time = -0.18 m/s, Step Duration = 0.06 s, Step Length = -0.07 m, Total Number of Steps = -28, and Total Distance Walked = -24 m.

**Conclusion:** These MCIDs provide a framework for interpreting meaningful gait changes and integrating digital measures into MS outcome evaluation. Beyond facilitating novel clinical trial endpoints to evaluate treatment efficacy, they enable objective, real-world monitoring to advance personalized patient care.

# 1 Introduction

Multiple sclerosis (MS) is a chronic neurological disorder characterized by inflammation, demyelination, and neurodegeneration.[1] Gait impairments are among the most prevalent and impactful manifestations of MS, affecting patients' quality of life, functional independence, and overall well-being.[1–4] Accurately assessing and quantifying changes in gait is, therefore, critical for both monitoring disease progression and evaluating the efficacy of therapeutic interventions.

Advances in digital health technologies, particularly the use of smartphone-based inertial measurement units (IMUs), have enabled the collection of continuous and high-resolution gait data in both clinical and real-world settings.[5] Gait measures, such as total distance walked and spatiotemporal measures like step velocity, step duration, and step length derived from smartphone IMUs during remote walking tests offer a promising complement to traditional clinical assessments.[6] However, for these digital gait measures to be interpretable and actionable, it is critical to understand what constitutes a clinically meaningful change. The minimal clinically important difference (MCID) is the most common approach, and it represents the smallest within-patient change in an outcome measure that is meaningful to patients, reflecting a tangible improvement or worsening in their daily lives.[7] MCIDs are typically derived using anchor-based methods. Such approaches evaluate changes on a target scale by linking them to an external reference (or anchor) that captures clinical meaningfulness.[8]

Currently, there is limited published evidence on MCID values for digital gait measures, which limits their integration into medical research and clinical decision-making. To address this gap, we estimated MCIDs for smartphone-based digital gait measures using an anchor-based method[8] that included multiple clinician-administered and patient-reported outcomes. The analyses were conducted in a prospective cohort of patients with progressive MS enrolled into the CONSONANCE (NCT03523858) study over 2 years. Additionally, this work was supplemented with qualitative results to ensure these digital gait measures are meaningful and relevant to patients.

# 2 Methods

## 2.1 CONSONANCE

CONSONANCE is an ongoing phase 3b, multicenter, open-label, single-arm study investigating the effectiveness and safety of ocrelizumab in progressive MS (PMS).[9] Key inclusion criteria were: age 18–65 years, a confirmed diagnosis of progressive MS according to the 2010 McDonald diagnostic criteria for primary progressive MS[10] or the Lublin 2014 criteria for progressive MS[11], an EDSS score of ≤6.5, and documented disability progression independent of relapses over two years prior to enrollment. The first patient was enrolled on May 28, 2018.

The study design is illustrated in Figure 1A. Clinical study visits were scheduled every 24 weeks for up to 4 years, during which patients received ocrelizumab intravenously, and clinical assessments and patient-reported outcomes were collected. The study also includes continuous remote monitoring with a smartphone solution, consisting of multiple tests including a daily, unsupervised, smartphone-based Two-Minute Walk Test (2MWT). Data collected up to Week 96 were used for the MCID analyses.

The study protocol (ClinicalTrials.gov identifier number: NCT03523858) was approved by the relevant institutional review boards (IRBs)/ ethics committees (ECs). All participants provided written informed consent prior to any study procedures.

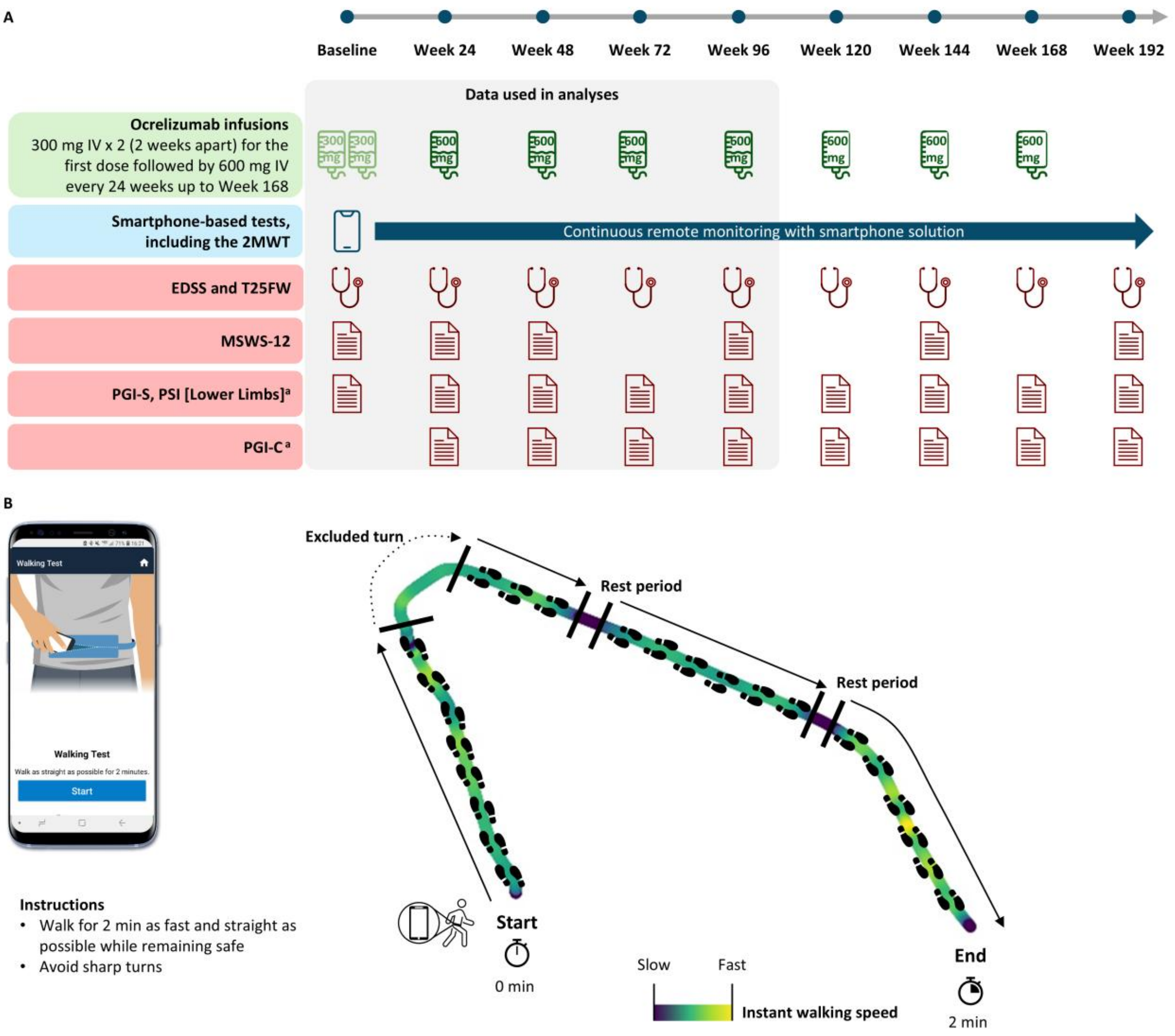

C

| Digital gait measure | Definition |
|---|---|
| Measures of gait quality | |
| Step Velocity, m/s | The median walking speed across all steps during a 2MWT during walking periods excluding sharp turns. |
| Step Velocity Scaled to Walking Time, m/s | A normalized measure of walking speed (Step Velocity, above) that accounts for the proportion of ambulation during a 2MWT. This measure adjusts for variability in rest behavior, as two individuals may have similar median velocities but differ in the proportion of time spent walking versus resting. |
| Step Duration, s | The median time taken for a single step during a 2MWT, reflecting the temporal rhythm of walking. |
| Step Length, m | The median distance covered per step during 2MWT, capturing the spatial aspect of gait. |
| Measures of gait quantity | |
| Total Number of Steps | The total count of steps taken during the 2MWT. |
| Total Distance Walked, m | The total distance covered during the 2MWT, calculated as the cumulative step length over all steps. |

**Figure 1.** Study design and the digital 2MWT. (A) The CONSONANCE study included daily digital 2MWT (blue) performed with a smartphone remotely in a home environment as well as performance-based, clinician-rated, and patient-reported measures used as anchors (red). (B) For the 2MWT, participants were instructed to walk as fast as possible but safely while avoiding sharp turns. Walking aid and/or orthotics were permitted, and patients could rest if needed. (C) Digital gait measures derived from the 2MWT include those assessing the quality of gait and those assessing the quantity of gait.

[a]Findings obtained with PGIC, PGIS, and PSI [Lower Limbs] are presented in the supplementary appendix.

[b]Note that not all assessments were administered at each study visit.
2MWT, Two-Minute Walk Test; EDSS, Expanded Disability Status Scale; IV, intravenous; MSWS-12, 12-item Multiple Sclerosis Walking Scale; PGI-C, Patient Global Impression of Change; PGI-S, Patient Global Impression of Severity; PPMS, primary progressive multiple sclerosis; PSI [Lower Limbs], Patient Impression of Severity for Lower Limbs; SPMS, secondary progressive multiple sclerosis; T25FW, Timed 25-Foot Walk.

## 2.2 Qualitative study

A separate qualitative study was conducted with patients with MS, which consisted of concept elicitation and cognitive debriefing interviews. All participants provided informed consent prior to interviews.

Individual concept elicitation interviews were conducted over the phone using a semi-structured interview guide to directly elicit information on MS symptoms of importance from a patient perspective. These interviews typically lasted 60 minutes. Participants were asked to report their most burdensome MS-related symptoms and how these symptoms impacted their daily lives. Subsequently, participants were specifically asked about their symptoms and impacts specific to their upper and lower limb function, cognitive function, and vision. Insights gathered from these interviews were used to establish the patient relevance of the digital gait measures.

In contrast, cognitive debriefing interviews were conducted in person or virtually via a secure online meeting platform (after the World Health Organization declared the global COVID-19 pandemic) and lasted approximately 90 minutes. After reviewing and performing the 2MWT, participants were asked about their perceived difficulty in performing the 2MWT, the MS-related symptoms that affect their test performance, the relevance of the perceived difficulty in performing the 2MWT to daily activities, and the extent to which the 2MWT captured important gait-related symptoms of MS. These cognitive debriefing interviews provided insights into which daily activities were affected by those MS-related symptoms that impacted the 2MWT performance, and were used to establish the relevance of the 2MWT in capturing gait-related MS symptoms that impact the daily lives of MS patients.

## The 2-Minute Walk Test (2MWT) and derived digital gait measures

Patients performed a daily, remote, walking test while carrying a provisioned Samsung Galaxy J7 smartphone around the waist in a belt bag (Figure 1B). Participants were instructed to walk as fast as possible, but safely, while avoiding sharp turns for 2 minutes. They were allowed to use their usual walking aid and/or orthotic or rest if needed during the 2MWT.

Sensor data were recorded during each 2MWT with inertial measurement units (IMUs) embedded in the smartphone, consisting of a triaxial accelerometer and a triaxial gyroscope (sampling frequency: 50 Hz). The IMU data were processed to detect steps using a validated gait analysis pipeline capable of identifying and removing turns and rests while detecting key gait events such as toe-off and heel strike events with high accuracy.[12] From these gait events, a wide range of digital gait measures were derived that capture both the quality (how did the patient walk; i.e., Step Velocity, Step Velocity Scaled to Walking Time, Step Duration, or Step Length) and quantity of gait (how much did they walk in 2 minutes; i.e., Total Number of Steps or Total Distance Walked). These gait measures are defined in Figure 1C and have previously undergone technical and clinical validation.[13–15]

## 2.3 Anchors for estimating MCIDs

Multiple anchors were administered during study visits, unlike the digital gait measures that were collected remotely in a real-life environment (Figure 1A). These include performance-based measures (Timed 25-Foot Walk [T25FW]),[16] clinician-rated measures (EDSS and its Ambulation Score subcomponent),[17,18] and patient-reported outcome measures (12-item Multiple Sclerosis Walking Scale [MSWS-12])[19].

Additional anchors include patient-reported outcomes on lower-limb specific symptoms (Patient Impression of Severity for Lower Limbs [PSI [Lower Limbs]]), disease severity (Patient Global Impression of Severity [PGI-S]), and change in disease severity (Patient Global Impression of Change [PGI-C]). These anchors represent scales specifically developed for the CONSONANCE study and have not been fully validated. Hence, findings obtained with these anchors are reported in the Supplementary Appendix.

**Table 1. Anchors used to estimate MCID values.**

| Anchor | Concept of interest | Definition | Anchor-defined thresholds based on change from baseline at Week 96 | |
|---|---|---|---|---|
| | | | **Worsening group** | **Stable group** |
| Performance-based measure | | | | |
| T25FW[20] | Gait capacity | Time needed to walk 25 feet as quickly and safely as possible | ≥20% increase[21] | >-20% to <20% change |
| Clinician-rated measures | | | | |
| Ambulation Score[18,22] | Overall walking and ambulation impairment | A subscore of the EDSS that is based on the distance a patient can walk, the need for walking aids, and the presence of other mobility limitations (range: 0 [= unrestricted ambulation] to 16 [= death due to MS]) | ≥1-point increase | 0-point change |
| EDSS[18,22,23] | Overall MS-related disability, with an emphasis on gait | A standardized scale from 0 (= normal neurological exam) to 10 (= death due to MS) that evaluates 8 different functional systems and the patient's level of mobility | ≥1-point increase if baseline EDSS ≤5.5; ≥0.5-point increase if baseline EDSS = 6.0 or 6.5 | >-1 to <1-point change if baseline EDSS ≤5.5; >-0.5 to <0.5-point change if baseline EDSS = 6.0 or 6.5 |
| Patient-reported measure | | | | |
| MSWS-12[19] | Impact of MS on walking ability | Total score across 12 items that evaluate the impact of MS on different aspects of walking, including balance, fatigue, and difficulty in navigating stairs (range transformed score: 0 [= no impairment] to 100 [= highest possible impairment]) | ≥8-point increase[24] | >-8 to <8-point change |

EDSS, Expanded Disability Status Scale; MSWS-12, 12-item Multiple Sclerosis Walking Scale; T25FW, Timed 25-Foot Walk.

## 2.4 Computation of changes from baseline to Week 96

For digital gait measures, longitudinal change was determined using median values aggregated over 4-week windows at baseline and Week 96. In contrast, clinical anchors were assessed via predefined absolute step changes from baseline at Week 96 (e.g., ≥1-point increase on EDSS for patients with a baseline EDSS of ≤5.5 or ≥0.5-point increase on EDSS for patients with a baseline EDSS of 6.0 or 6.5), except for the T25FW, where percent change was calculated to align with the established ≥20% worsening threshold[21] (Table 1). For each anchor, patients were dichotomized into "worsened" or "stable" cohorts based on predefined 96-week change thresholds (Table 1). Consequently, patient classification was anchor-specific, reflecting the distinct aspects of mobility captured by each metric.

## 2.5 Statistical analysis

The analysis comprised multiple steps. First, data were described by (I) examining the distribution of the anchor scores at baseline, (II) visualizing the median change in the digital measures over the 96-week follow-up period, and (III) correlating the changes from baseline at Week 96 on the digital measures with corresponding changes on the anchors using Spearman rank-order correlation.

Next, MCID values were derived through the following procedure: (I) Group stratification: Split into aforementioned worsening groups (defined by anchor-specific progression thresholds) and stable groups (representing change below the threshold for clinical meaningfulness). (II) Change quantification: Median changes were calculated for both groups. Following recommendations for the computation of MCIDs[25,26], the median changes on the digital gait measures in the worsening group correspond to the 50$^{th}$ percentile of the Empirical Cumulative Distribution Function (ECDF) of the respective anchor. Separate ECDFs were generated for each gait measure–anchor pairing. (III) Group comparison: Differences between the worsening and stable groups were examined through visual inspection and Mann-Whitney U test, with changes in the stable group expected to be minimal. (IV) Triangulation: To derive a unified MCID for each digital measure, individual median changes in the worsening groups across different anchors were synthesized using the correlation-weighted approach of Trigg and Griffiths 2021.[27] This approach uses a meta-analytic framework to assign weights to different anchor-based MCIDs, determined by the strength of Fisher's z-transformed correlations (anchors with higher correlations are considered more reliable and are thus given greater weight in the calculation).

All analyses were conducted using Python, leveraging libraries such as `pandas` and `numpy` for data manipulation, `scipy` for statistical analyses, and `matplotlib` for ECDF visualization.

# 3 Results

## 3.1 Baseline demographics and disease characteristics

A total of 243 patients with progressive MS had digital gait data available at both baseline and Week 96, and were included in the MCID analysis. Baseline data as described by the clinical anchors show a cohort with high disability accrual (mean EDSS of 4.8 and 42% of patients having an EDSS score of 6.0 or 6.5), including significant walking impairment (mean T25FW time of 11.9 seconds). MSWS-12 values (mean: 58.2) suggest moderate-to-severe self-reported walking limitations. Additional baseline demographics and disease characteristics are described in Table 2.

**Table 2. Baseline demographics and disease characteristics for MCID analysis.**

| Variable | N = 243 |
|---|---|
| Female, n (%) | 125 (51.4) |
| Age, years, mean (SD) | 49.3 (9.3) |
| Type of MS, n (%) | |
| PPMS | 111 (45.7) |
| SPMS | 132 (54.3) |
| Time since MS diagnosis (years), mean (SD) | 7.7 (7.6) |
| Time since MS symptom onset (years), mean (SD) | 11.1 (7.5) |
| EDSS | |
| Mean (SD) | 4.8 (1.4) |
| 1.5 – 3.5, n (%) | 65 (26.7) |
| 4 – 5.5, n (%) | 75 (30.9) |
| 6.0 – 6.5, n (%) | 103 (42.4) |
| Ambulation Score, mean (SD) | 3.7 (2.9) |
| T25FW, seconds, mean (SD) | 11.9 (10.3) |
| MSWS-12 version 2 transformed, total score, mean (SD) | 58.2 (24.4) |

BMI, body mass index; EDSS, Expanded Disability Status Scale; MSWS-12, 12-item Multiple Sclerosis Walking Scale; PPMS, primary progressive multiple sclerosis; SPMS, secondary progressive multiple sclerosis; T25FW, Timed 25-Foot Walk.

## 3.2 Digital gait measures show worsening over time

All digital gait measures showed worsening from baseline up to Week 96 (Supplementary Figure S1). Gait speed decreased, with patients showing a gait pattern with shorter steps (Step Length) while taking longer time between steps (Step Duration). The number of completed steps and distance walked during the 2MWT also decreased over time.

## 3.3 Longitudinal correlations confirm suitability of anchors

Longitudinal Spearman rank-order correlations were in the expected direction, with worsening on the anchors associated with worsening in all digital gait measures (Table 3). Overall, change in patient-perceived walking impairment as measured by MSWS-12 demonstrated the strongest correlations with digital gait measures, followed by the T25FW; Ambulation Score showed the weakest correlations. For multiple digital gait measures, at least one correlation with $\rho \geq 0.3$ was observed, with the highest absolute correlation between Step Velocity Scaled to Walking Time and MSWS-12 ($\rho = -0.33$). The magnitude of the correlations for each anchor was highly consistent for most digital gait measures, except for Step Duration which showed weaker correlations ($\rho < 0.2$) across all anchors.

**Table 3: Longitudinal Spearman rank-order correlations and median changes from baseline at Week 96 on digital gait measures by anchor.**

| Digital gait measure/ anchor | Longitudinal Spearman rank-order correlations[a] | | | Median change from baseline at Week 96 | | | | | |
|---|---|---|---|---|---|---|---|---|---|
| | | | | Worsening groups | | Stable groups | | | |
| | n | ρ | p-value[b] | n | Median change | n | Median change | Ratio[c] | p-value[d] |
| Step Velocity, m/s | | | | | | | | | |
| T25FW | 225 | -0.223 | <0.001 | 71 | -0.134 | 114 | -0.041 | 3.3 | <0.001 |
| Ambulation Score | 232 | -0.166 | 0.011 | 52 | -0.145 | 146 | -0.051 | 2.9 | 0.054 |
| EDSS | 233 | -0.218 | <0.001 | 33 | -0.189 | 171 | -0.049 | 3.9 | 0.002 |
| MSWS-12 | 234 | -0.311 | <0.001 | 69 | -0.162 | 72 | -0.025 | 6.5 | <0.001 |
| Step Velocity Scaled to Walking Time, m/s | | | | | | | | | |
| T25FW | 225 | -0.255 | <0.001 | 71 | -0.158 | 114 | -0.041 | 3.8 | <0.001 |
| Ambulation Score | 232 | -0.162 | 0.013 | 52 | -0.156 | 146 | -0.053 | 2.9 | 0.054 |
| EDSS | 233 | -0.209 | 0.001 | 33 | -0.218 | 171 | -0.053 | 4.1 | 0.002 |
| MSWS-12 | 234 | -0.330 | <0.001 | 69 | -0.171 | 72 | -0.052 | 3.3 | <0.001 |
| Step Duration, | | | | | | | | | |
| s | | | | | | | | | |
| T25FW | 225 | 0.137 | 0.040 | 71 | 0.06 | 114 | 0.02 | 3.0 | 0.007 |
| Ambulation Score | 232 | 0.132 | 0.045 | 52 | 0.04 | 146 | 0.02 | 2.3 | 0.431 |
| EDSS | 233 | 0.156 | 0.017 | 33 | 0.06 | 171 | 0.02 | 3.0 | 0.095 |
| MSWS-12 | 234 | 0.192 | 0.003 | 69 | 0.06 | 72 | 0.02 | 3.0 | 0.007 |
| Step Length, m | | | | | | | | | |
| T25FW | 225 | -0.280 | <0.001 | 71 | -0.062 | 114 | -0.010 | 6.1 | <0.001 |
| Ambulation Score | 232 | -0.164 | 0.012 | 52 | -0.065 | 146 | -0.017 | 3.9 | 0.013 |
| EDSS | 233 | -0.185 | 0.005 | 33 | -0.081 | 171 | -0.017 | 4.8 | <0.001 |
| MSWS-12 | 234 | -0.310 | <0.001 | 69 | -0.066 | 72 | -0.009 | 7.8 | <0.001 |
| Total Number of Steps | | | | | | | | | |
| T25FW | 225 | -0.235 | <0.001 | 71 | -31.5 | 114 | -11.0 | 2.9 | <0.001 |
| Ambulation Score | 232 | -0.180 | 0.006 | 52 | -27.5 | 146 | -13.0 | 2.1 | 0.053 |
| EDSS | 233 | -0.157 | 0.017 | 33 | -32.5 | 171 | -13.0 | 2.5 | 0.004 |
| MSWS-12 | 234 | -0.226 | <0.001 | 69 | -22.0 | 72 | -16.3 | 1.4 | 0.085 |
| Total Distance Walked, m | | | | | | | | | |
| T25FW | 225 | -0.232 | <0.001 | 71 | -19.0 | 114 | -6.6 | 2.9 | <0.001 |
| Ambulation Score | 232 | -0.161 | 0.014 | 52 | -25.9 | 146 | -9.1 | 2.9 | 0.051 |
| EDSS | 233 | -0.188 | 0.004 | 33 | -28.9 | 171 | -9.1 | 3.2 | 0.004 |
| MSWS-12 | 234 | -0.317 | <0.001 | 69 | -24.6 | 72 | -9.5 | 2.6 | 0.003 |

[a]Longitudinal Spearman rank-order correlations were computed between median changes from baseline at Week 96 on digital gait measures and corresponding changes on the anchors across all participants.
[b]P-values were computed using the SciPy package for Python.
[c]The ratio was computed as the median change of the worsening group divided by the median change of the stable group.
[d]P-values were computed with the Mann-Whitney U test.
EDSS, Expanded Disability Status Scale; MSWS-12, 12-item Multiple Sclerosis Walking Scale; T25FW, Timed 25-Foot Walk.

## 3.4 Median changes from baseline are more than twice as large in the worsening groups than in the stable groups

The distribution of changes from baseline at Week 96 on the digital gait measures are depicted in the ECDF plots (Figure 2), while median changes from baseline are reported in Table 3. In the worsening group, median changes were in the expected direction and consistent across all anchors. Furthermore, these changes were at least twice as large as corresponding median changes from baseline in the stable group (Table 3). This increases the confidence that the median changes from baseline observed in the worsening group reflect a meaningful change.

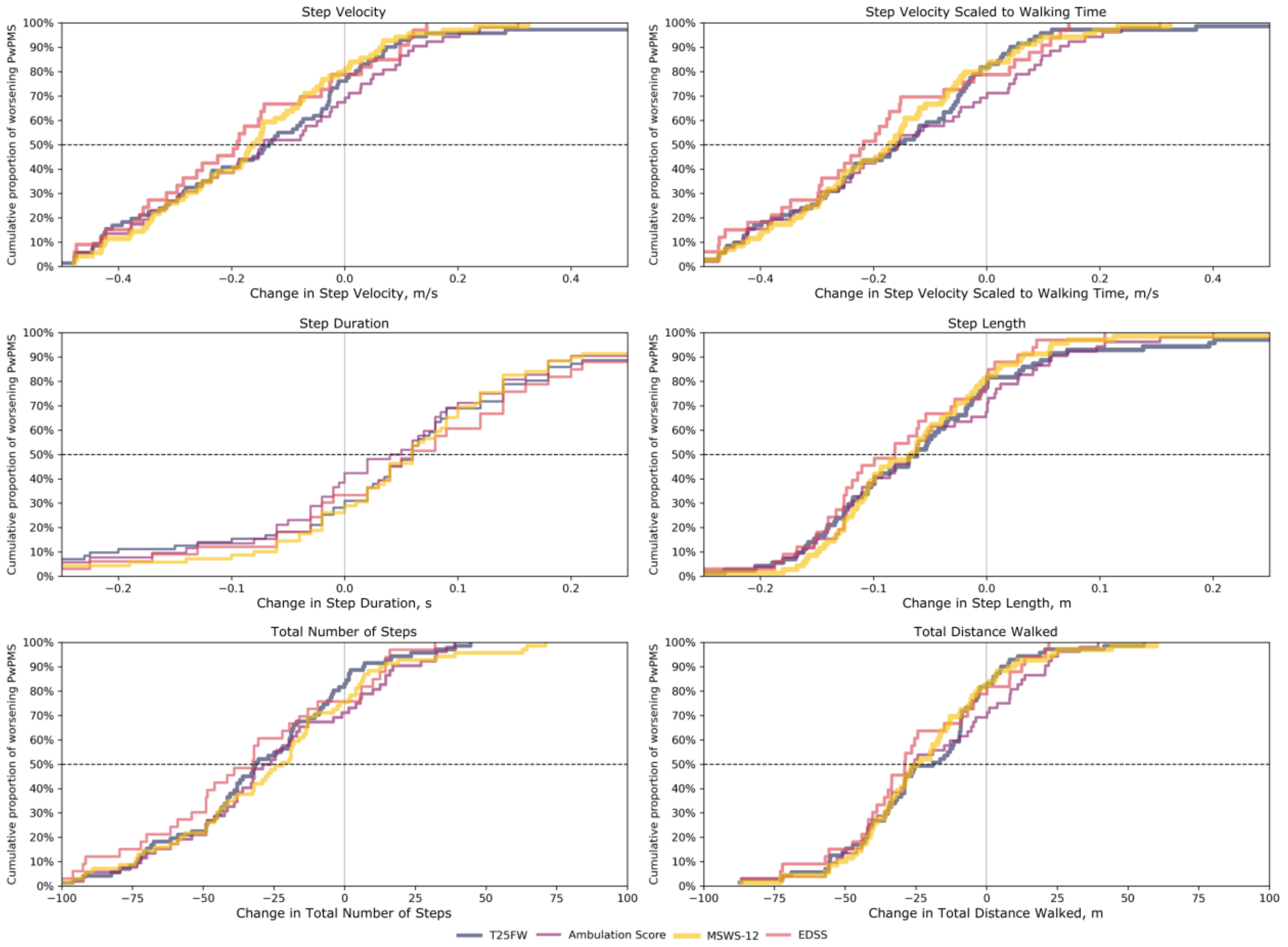

**Figure 2.** Empirical cumulative distribution functions (ECDFs) for changes on digital gait measures from baseline at Week 96 are shown for anchor-specific worsening groups. Each curve represents the subset of patients who met the respective anchor-defined threshold for meaningful worsening reported in Table 1. The line width of each ECDF corresponds to the strength of the Spearman's rho longitudinal correlation between the digital measure and the respective anchor reported in Table 3. EDSS, Expanded Disability Status Scale; MSWS-12, 12-item Multiple Sclerosis Walking Scale; T25FW, Timed 25-Foot Walk.

## 3.5 MCID estimates of MS disease progression

Triangulation of the median changes from baseline obtained with the different anchors yielded the following MCIDs of MS disease progression (Table 4): -0.16 m/s for Step Velocity, -0.18 m/s for Step Velocity Scaled to Walking Time, 0.06 s for Step Duration, -0.07 m for Step Length, -28 for Total Number of Steps, and -24 m for Total Distance Walked.

**Table 4. MCID estimates of digital gait measures.**

| Digital gait measures | MCID estimate |
|---|---|
| Step Velocity, m/s | -0.16 |
| Step Velocity Scaled to Walking Time, m/s | -0.18 |
| Step Duration, s | 0.06 |
| Step Length, m | -0.07 |
| Total Number of Steps | -28 |
| Total Distance Walked, m | -24 |

## 3.6 The 2MWT is a gait assessment highly relevant to patients

A total of 40 patients agreed to participate in the concept elicitation interviews, and 16 patients in the cognitive debriefing interviews. Their demographic and disease characteristics are reported in Supplementary Table S4. Compared to the MCID cohort, the qualitative study enrolled a greater proportion of female participants (concept elicitation cohort: 62.5%; cognitive debriefing cohort: 68.7% vs MCID cohort: 51.4%) and also patients with relapsing-remitting MS. Patient-reported EDSS scores were, however, comparable to the clinician-rated EDSS of the MCID cohort (concept elicitation cohort: 5.0 [SD: 1.9]; cognitive debriefing cohort: 4.7 [2.2]; MCID cohort: 4.8 [1.4]). This suggests that the level of MS-related disability was similar across both cohorts.

Several patients reported that MS hindered their ability to walk fast, with many noting a significant transition to a slower pace or the required use of assistive devices, stating *"I have to walk a lot slower"*, *"The fact that I can't walk normal. I can't walk with any quickness or anything, any speed."*, or *"I used to walk 10-minute miles, and now it's extremely slow."* Such reductions in walking speed are directly measured with Step Velocity, Step Velocity Scaled to Walking Time, and Step Duration. Beyond walking speed, patients also highlighted fatigue-related interruptions necessitating breaks while walking as a barrier to mobility (*"I have to take breaks after a few moments"*). Additionally, gait impairments, such as difficulty moving their legs, that result in shorter steps—described by one patient as *"It's just that sometimes I have trouble moving my legs, it's just as simple as that, sometimes it's a short taking steps […]"*—can be quantified with Step Length. A broader decline in endurance and stamina also emerged as a recurring theme, with patients noting *"I can't walk as far as I used to"*, *"Now I used to take like long walks. Now I can't."*, or *"Then it got to where I could only walk a short distance […]"*. Reduced endurance and stamina directly translate to a lower Total Number of Steps and Total Distance Walked. Detailed patient statements are provided in Supplementary Table S5.

The patient relevance of the digital measures is further underscored by the patients' description of how the symptoms impacting their performance on the 2MWT also impede daily life activities. Specifically, walking and completing their every-day routines were identified as daily activities impacted by the same MS symptoms that affect their 2MWT performance (*"walking in the store"*, *"walking into church or walking around the house"*, *"vacuuming"*, *"walking, going places and doing things"*). Collectively, these findings suggest that the digital measures are highly relevant to patients with MS. Detailed patient statements are provided in Supplementary Table S6.

# 4 Discussion

This study determined MCIDs of progression in MS for patient-relevant, smartphone-derived digital gait measures that capture both the quality and quantity of gait. These MCIDs were estimated through a method leveraging multiple established, performance-based, clinician-rated, and patient-reported anchors. This study elevates the utility of these digital gait measures, enabling the development of novel, more sensitive progression-based endpoints for MS clinical trials. Furthermore, it strengthens the clinical applicability of these measures by providing a framework for interpreting remote monitoring data in clinical practice.

## 4.1 MCID estimates

The proposed MCID estimates for MS disease progression (Step Velocity: -0.16 m/s; Step Velocity Scaled to Walking Time: -0.18 m/s; Step Duration: 0.06 s; Step Length: -0.07 m; Total Number of Steps: -28; Total Distance Walked: -24 m) are consistent with prior research. Reduced walking speed is a hallmark of disease progression and has been shown to decrease by approximately 0.12 m/s for every 1-point increase in EDSS.[28] Our MCID estimate for Step Velocity (-0.16 m/s) is comparable despite using a lower threshold of a 0.5-point increase on the EDSS in patients with a baseline EDSS of 6.0–6.5. Similarly, Step Length and Step Duration have been reported to decrease by 0.04 m and increase by 0.04 s per 1-point EDSS increase,[28] aligning with our slightly larger MCID estimates of -0.07 m and 0.06 s, respectively. It should be noted that these findings from the literature have been derived from cross-sectional comparisons between patients at different EDSS levels whereas our MCID estimates were based on longitudinal changes within patients. Similarly, our MCID estimate for Total Distance Walked is comparable to previously reported MCIDs obtained with T25FW (31.9 m) and EDSS (29.7 m) as anchors.[29] Overall, the similarity of MCID estimates when compared to literature reinforces the validity of these measures in quantifying meaningful changes in gait performance.

## 4.2 A conservative approach to estimating MCIDs

The proposed MCIDs are robust and conservative estimates for four main reasons. First, the anchor-based method used here is generally more accepted than distribution-based methods.[8] It is also more conservative than other anchor-based methods such as the receiver-operating-characteristic–based (ROC-based) method.[29] The ROC-based method finds the threshold that best discriminates between the worsening and stable group. While this approach optimizes the trade-off between sensitivity and specificity, the resulting MCID will be smaller than with MCID estimates based on the median change from baseline in the worsening group, and thus less conservative.[30] Moreover, the use of multiple relevant anchors—including performance-based, clinician-rated, and patient-reported outcome measures—along with progression-based thresholds increases the confidence that the obtained MCID estimates capture meaningful clinical progression, in particular in light of the consistency in the estimates across the multiple anchors. Second, the CONSONANCE dataset was sufficiently large for deriving robust MCID estimates. Third, by focusing on the distribution of change of the worsening group to establish MCIDs, the analysis minimizes potential biases introduced by changes in the stable group. For example, we observed that the median change from baseline in the stable group did not perfectly center around zero. It is, therefore, plausible that some patients classified as stable according to the anchors may exhibit subclinical or not-yet-meaningful changes in digital gait measures. While this highlights the sensitivity of digital gait measures, it also underscores the need for caution when interpreting small deviations in the stable groups. Thus, computing the MCID estimates based on median changes from baseline in the worsening group ensures that the MCIDs reflect a meaningful change from zero. These median changes were comparable to those obtained with supplementary anchors such as PSI [Lower Limbs], PGI-S, and PGI-C (Supplementary Appendix), which reinforces the robustness of the MCID estimates.

## 4.3 Implications for clinical trial and practice

The digital gait measures probe multiple aspects of gait that are relevant to people with MS and have shown sensitivity to longitudinal change.[15] We expanded on these previously reported findings by deriving MCID estimates that provide means for interpreting meaningful changes on the remote, smartphone-based 2MWT. These estimates, therefore, enable longitudinal assessments of the patient's daily functioning of gait as opposed to the evaluation of gait in a supervised, clinical setting.

In clinical trials, these MCID estimates enable the identification of clinically meaningful progression of gait. They may also serve as thresholds for time-to-event-based endpoints, such as time to gait progression, which are increasingly being used in trials of disease-modifying therapies (DMTs).[31] In clinical practice, remote, digital tests performed in between clinic visits can complement clinical assessments by providing high-frequency, objective, rater-independent monitoring of walking ability.[32] Thus, the proposed MCIDs provide a framework for interpreting patient-level changes in gait measures and understanding the impact of disease progression or treatment effects.

## 4.4 Limitations

These analyses come with certain limitations. First, due to sample size constraints, a further split of the worsening group into subcategories (e.g., mild worsening vs. more severe worsening) was not appropriate, which could have improved the precision of the MCID estimates or the understanding of their robustness to baseline invariance. Second, while differential treatment effects can be excluded as all patients received ocrelizumab, the generalizability of these MCIDs is limited to the specific cohort studied (a progressive MS cohort that did not include patients with low levels of MS-related disability). Nonetheless, the similarity between our MCID estimates and those described in the literature for comparable digital gait measures suggests that our estimates generalize well to other MS populations. Third, the low correlations between some digital gait measures and anchors may lead to underestimation of MCID values.[33] However, the correlation-weighted triangulation method should mitigate any such potential bias.

Future research should explore whether the MCIDs remain consistent across varying baseline levels of digital gait measures. Validation in larger, independent MS cohorts, including natural history studies or investigations of other disease-modifying therapies (DMTs), will be critical to further confirming the robustness of the MCID estimates and their generalizability. Additionally, while this study focused on MCIDs for progression due to the progressive nature of MS, future work could explore MCIDs for improvement, which may be relevant in specific clinical contexts.

# 5 Conclusion

This study establishes robust and clinically meaningful MCIDs for smartphone-derived digital gait measures in progressive MS, leveraging a comprehensive anchor-based methodology and diverse clinical, patient-reported, and performance-based measures. The following suggested MCID reference points for Step Velocity (-0.16 m/s), Step Velocity Scaled to Walking Time (-0.18 m/s), Step Duration (0.06 s), Step Length (-0.07 m), Total Number of Steps (-28), and Total Distance Walked (-24 m) provide a conservative yet reliable framework for interpreting gait progression in people with multiple sclerosis. These thresholds have important implications for clinical trials, enabling the evaluation of disease-modifying therapies, and for real-world monitoring, offering objective insights into gait changes. The findings highlight the sensitivity and clinical relevance of digital gait measures, while acknowledging the need for further validation in larger, independent MS populations. These MCIDs lay the groundwork for integrating digital gait measures into the evolving landscape of MS outcome measures, advancing personalized care and trial design.

# Declarations

## Data Availability

Up-to-date details on Roche's Global Policy on Sharing of Clinical Study Information and how to request access to related clinical study documents are available via https://go.roche.com/data_sharing. Request for the data underlying this publication requires a detailed, hypothesis-driven, statistical analysis plan that is collaboratively developed by the requestor and company subject matter experts. Such requests should be directed to dbm.datarequest@roche.com for consideration. Anonymized records for individual patients across more than one data source external to Roche cannot, and should not, be linked due to a potential increase in risk of patient reidentification.

## Competing Interests

*MDR* is a contractor for F. Hoffmann-La Roche Ltd.
*BF* is a consultant for F. Hoffmann-La Roche Ltd.
*DS* is a consultant for F. Hoffmann-La Roche Ltd.
*CS* is an employee of F. Hoffmann-La Roche Ltd.
*EB* is an employee of and shareholder in F. Hoffmann-La Roche Ltd.

*LL* has received compensation for consulting services from Almirall, EXCEMED, F. Hoffmann-La Roche Ltd, Merck, Novartis, Biogen, Janssen-Cilag and Bristol Myers Squibb.
*AK* is an employee of and shareholder in F. Hoffmann-La Roche Ltd.
*GC* has received compensation for participation in data and safety monitoring boards for Applied Therapeutics, AI Therapeutics, AMO Pharma, Argenx, AstraZeneca, Avexis Pharmaceuticals, Bristol Myers Squibb, CSL Behring, Cynata Therapeutics, DiaMedica Therapeutics, Horizon Pharmaceuticals, Immunic, Inhibrix, Karuna Therapeutics, Kezar Life Sciences, Medtronic, Merck, Meiji Seika Pharma, Mitsubishi Tanabe Pharma Holdings, Prothena Biosciences, Novartis, Pipeline Therapeutics (Contineum), Regeneron, Sanofi-Aventis, Teva Pharmaceuticals, United BioSource, and University of Texas Southwestern, and in consulting for Alexion, Antisense Therapeutics–Percheron, Avotres, Biogen, Clene Nanomedicine, Clinical Trial Solutions, Endra Life Sciences, Cognito Therapeutics, F. Hoffman-La Roche, Genentech, Genzyme, Hoya Corporation, Immunic, Immunosis Pty, Klein-Buendel Incorporated, Kyverna Therapeutics, Linical, Merck–Serono, Noema, Neurogenesis, Perception Neurosciences, Protalix Biotherapeutics, Regeneron, Revelstone Consulting, SAB Biotherapeutics, Sapience Therapeutics, Scott & Scott, and Tenmile; has received institutional grants from University of Alabama at Birmingham, AL, USA; and holds non-financial leadership or fiduciary roles in the Birmingham Jewish Federation, Birmingham Jewish Foundation, and Graffman Endowment Committee.
*HB* is an employee of Monash University and board member and managing director of MSBase Foundation. He has received conference travel compensation from Novartis and Merck. His institution received honoraria for talks and lectures from Merck, F. Hoffmann-La Roche Ltd, Novartis. He has received personal compensation for talks from Neuraxpharm. His institution has received compensation for activities in study steering committees, advisory boards, and discussion forums from F. Hoffmann-La Roche Ltd, Novartis, and Merck and research grants from F. Hoffmann-La Roche Ltd., Argenx, Novartis, Merck, UCB Pharma, the Medical Research Future Fund Australia, NHMRC Australia, Trish Foundation, MS Australia, and Monash University.
*LC* is an employee of and shareholder in F. Hoffmann-La Roche Ltd.

## Funding

This research was funded by F. Hoffmann-La Roche Ltd, Basel, Switzerland.

## CRediT authorship contribution statement

*MDR*: Conceptualization, Data Curation, Formal Analysis, Methodology, Project Administration, Software, Supervision, Visualization, Writing – Original Draft Preparation, Writing – Review & Editing
*BF*: Conceptualization, Formal Analysis, Methodology, Visualization, Writing – Original Draft Preparation, Writing – Review & Editing
*DS*: Data Curation, Software, Formal Analysis, Writing – Review & Editing
*CS*: Data Curation, Software, Formal Analysis, Writing – Review & Editing
*EB*: Writing – Review & Editing
*LL*: Conceptualization, Investigation, Writing – Review & Editing
*AK*: Supervision, Project Administration, Writing – Review & Editing
*GC:* Writing – Review & Editing
*HB*: Conceptualization, Investigation, Writing – Review & Editing
*LC*: Funding Acquisition, Writing – Review & Editing

## Acknowledgements

The authors would like to thank all study participants and their families. Editorial and writing support was provided by Sven Holm, PhD, contractor for F. Hoffmann-La Roche Ltd. The authors confirm that all intellectual content, data interpretation, and conclusions were generated by the authors.

# *Supplementary appendix to:* Establishing the Minimal Clinically Important Difference (MCID) for Smartphone-Derived Gait Measures in Multiple Sclerosis

Mike D Rinderknecht,[a] Bernhard Fehlmann,[a] Dimitar Stanev,[a] Cedric Simillion,[a] Ernst Bos,[a] Letizia Leocani,[b,c] Agne Kazlauskaite,[a] Gary Cutter,[d] Helmut Butzkueven,[e,f] Licinio Craveiro[a]

[a] F. Hoffmann-La Roche Ltd., Basel, Switzerland
[b] Faculty of Medicine, University Vita-Salute San Raffaele, Milan, Italy
[c] Scientific Institute IRCCS San Raffaele, Milan, Italy
[d] Department of Biostatistics, University of Alabama at Birmingham, Birmingham, AL, USA
[e] Department of Neuroscience, School for Translational Medicine, Monash University, Melbourne, Australia
[f] Department of Neurology, Alfred Health, Melbourne, Australia

**Corresponding author**
Mike D Rinderknecht
mike.rinderknecht@external.roche.com

# Methods

## Additional anchors used in supplementary analyses

The Patient Impression of Severity for Lower Limbs (PSI [Lower Limbs]), Patient Global Impression of Severity (PGI-S), and Patient Global Impression of Change (PGI-C) are three patient-reported outcomes developed by Roche. The PSI [Lower Limbs] asks patients to rate the severity of multiple sclerosis (MS)-related symptoms affecting their lower limbs over the past 7 days on a 5-point Likert scale. The PGI-S is a similar scale but asks the patients the overall severity of their MS-related symptoms, including those affecting their hands and arms, lower limbs, and cognition. Response options on both the PSI [Lower Limbs] and PGI-S range from “none” to “very severe”. In contrast, the PGI-C asks the patients to describe their overall change in MS-related symptoms over the past 6 months on a 7-point Likert scale, with response options ranging from “very much better” to “very much worse”. As these three patient-reported outcomes have not been fully validated, they have been included in supplementary analyses. Their schedule of assessment is reported in Figure 1A.

Thresholds applied to these PSI, PGI-S, and PGI-C scales to define worsening and stable groups are reported in Supplementary Table S1. Subsequent changes from baseline on both digital gait measures and these anchors were computed in the same fashion as for the other anchors with one notable exception: as PGI-C has a recall period of 6 months (= 24 weeks), changes at Week 96 on digital gait measures were computed from baseline (“from BL”) and additionally also from Week 72 (“last 24w”) to match PGI-C’s recall period as illustrated in Supplementary Figure S2.

# Results

## Longitudinal correlations: supplementary analysis

Longitudinal correlations between digital gait measures and either PSI [Lower Limbs], PGI-S, or PGI-C were overall in the expected direction, with worse performance on the 2MWT being associated with more severe or greater worsening in MS-related symptoms (Supplementary Table S2). The only instances of counterintuitive directionality were correlations between the change from Week 72 at Week 96 in digital gait measures and PGI-C. However, in each of these instances, correlations were non-significant and near zero (correlation with Step

Duration: ρ = -0.004, p = 0.964; Total Number of Steps: ρ = 0.003, p = 0.975; Total Distance Walked: ρ = 0.034, p = 0.671; Supplementary Table S2).

## Median changes on digital gait measures in both worsening and stable groups: supplementary analysis

Median changes on digital gait measures were generally consistent with those used for deriving the MCID estimates (compare Table 3 with Supplementary Table S3). Notable exceptions included the worsening groups defined by PGI-C when computing the median changes on the digital gait measure from Week 72 rather than from baseline. ECDF plots are provided in Supplementary Figure S3.

# Discussion

Findings obtained with PSI [Lower Limbs], PGI-S, and PGI-C were generally consistent with those obtained with the other anchors. A few notable exceptions were, however, observed. These can be attributed to either the smaller sample size in the worsening group defined by PGI-C or the computing of changes on digital gait measures over a period of just 24 weeks (from Week 72 at Week 96) rather than over a period of two years (from baseline at Week 96). Nonetheless, these findings suggest that comparable MCID estimates would have been derived with the triangulation method if median changes on digital gait measures in worsening groups defined by the PSI, PGI-S and PGI-C scales had been included.

## Contextualizing PSI, PGI-S, and PGI-C

Unlike the gait-specific anchors such as the MSWS-12, both PGI-C and PGI-S are global measures capturing the overall impact of MS symptoms. Nonetheless, they both showed meaningful relationships with digital gait measures. This likely reflects the importance of gait impairment in MS, which is one of the most frequent and impactful functional impairments experienced by people with MS,[1–6] and people with MS may place significant weight on their walking ability when rating the severity of their MS symptoms captured on the PGI-S. Furthermore, for the observed range of baseline EDSS scores (range: 1.5–6.5; with 42% [103/243] of patients having baseline EDSS scores ≥6.0), gait impairment mediated through pyramidal dysfunction is among the primary contributors to MS disability progression.[7] However, results obtained with PGI-C and PGI-S were not always consistent with results obtained with the other, more gait-related anchors. This underscores the importance of evaluating and comparing results from multiple anchors when deriving MCID estimates with an anchor-based approach.

Interestingly, PGI-C correlations with changes in digital gait measures over the last 24 weeks (from Week 72 to Week 96) were weaker than corresponding correlations with changes in digital gait measures from Baseline. A likely explanation is that changes in digital gait measures measured over a shorter timeframe of 24 (i.e., change from Week 72 at Week 96) vs 96 weeks (i.e., change from baseline at Week 96) tend to be smaller and, therefore, more likely to be impacted by measurement noise relative to the signal. These two effects can both contribute to weaker correlation strength and potentially stronger underestimation of MCIDs.[8] Additionally, PGI-C scores may also relate to acute changes related to functional domains other than gait that are impacted by MS. It is also conceivable that people with MS report on the PGI-C their perceived progression since the start of the study (e.g., initiation of treatment being a more relevant timepoint in patients' lives) rather than the change over the last 6 months (i.e., the recall period of the PGI-C).

# Supplementary tables and figures

**Supplementary Table S1. Additional anchors used in supplementary analysis.**

| Anchor | Concept of interest | Measure | Anchor-defined thresholds at Week 96 | |
|---|---|---|---|---|
| | | | Worsening group | Stable group |
| PSI [Lower Limbs] | Symptoms affecting lower limbs | Single item that evaluates the severity of symptoms (weakness, stiffness, or muscle spasms) affecting lower limbs (range: 1 = none to 5 = very severe) | ≥1-point worsening from baseline | >-1 to <1-point change from baseline |
| PGI-S | Global disease severity | Single item that evaluates the patient's current overall perception of MS disease severity (range: 1 = none to 5 = very severe) | ≥1-point worsening from baseline | >-1 to <1-point change from baseline |
| PGI-C | Change in disease severity | Single item that evaluates the patient's overall perception of the change within a recall period of 6 months (i.e., approximately 24 weeks) in MS disease severity (range: 1 = very much better to 7 = very much worse, with 4 = no change) | Score ≥5 (indicating worsening) | Score = 4 |

PGI-C, Patient Global Impression of Change; PGI-S, Patient Global Impression of Severity; PSI [Lower Limbs], Patient Global Impression of Severity for Lower Limbs.

**Supplementary Table S2. Longitudinal correlations between digital gait measures and PSI, PGI-S, and PGI-C scales.**

| Digital gait measure/ anchor | Change at Week 96 in digital gait measure | Spearman rank-order correlation | | |
|---|---|---|---|---|
| | | n | ρ | p-value |
| Step Velocity, m/s | | | | |
| PSI [Lower Limbs] | From BL | 47 | -0.118 | 0.428 |
| PGI-S | From BL | 47 | -0.327 | 0.025 |
| PGI-C | Last 24w | 158 | -0.084 | 0.294 |
| PGI-C | From BL | 175 | -0.277 | <0.001 |
| Step Velocity Scaled to Walking Time, m/s | | | | |
| PSI [Lower Limbs] | From BL | 47 | -0.122 | 0.412 |
| PGI-S | From BL | 47 | -0.302 | 0.036 |
| PGI-C | Last 24w | 158 | -0.090 | 0.259 |
| PGI-C | From BL | 175 | -0.280 | <0.001 |
| Step Duration, s | | | | |
| PSI [Lower Limbs] | From BL | 47 | 0.029 | 0.848 |
| PGI-S | From BL | 47 | 0.136 | 0.369 |
| PGI-C | Last 24w | 158 | -0.004 | 0.964 |
| PGI-C | From BL | 175 | 0.149 | 0.049 |
| Step Length, m | | | | |
| PSI [Lower Limbs] | From BL | 47 | -0.122 | 0.416 |
| PGI-S | From BL | 47 | -0.346 | 0.017 |
| PGI-C | Last 24w | 158 | -0.043 | 0.588 |
| PGI-C | From BL | 175 | -0.303 | <0.001 |
| Total Number of Steps | | | | |
| PSI [Lower Limbs] | From BL | 47 | -0.045 | 0.763 |
| PGI-S | From BL | 47 | -0.254 | 0.084 |
| PGI-C | Last 24w | 158 | 0.003 | 0.975 |
| PGI-C | From BL | 175 | -0.104 | 0.173 |
| Total Distance Walked, m | | | | |
| PSI [Lower Limbs] | From BL | 47 | -0.069 | 0.645 |
| PGI-S | From BL | 47 | -0.371 | 0.010 |
| PGI-C | Last 24w | 158 | 0.034 | 0.671 |
| PGI-C | From BL | 175 | -0.216 | 0.004 |

From BL, from baseline; Last 24, last 24 weeks (from Week 72); PGI-C, Patient Global Impression of Change; PGI-S, Patient Global Impression of Severity; PSI [Lower Limbs], Patient Global Impression of Severity for Lower Limbs.

**Supplementary Table S3. Median changes from baseline/ Week 72 at Week 96 on digital gait measures in subgroups defined by supplementary anchors.**

| Digital gait measure/ anchor | Change in digital gait measure | Worsening groups | | Stable groups | | Ratio | p-value |
|---|---|---|---|---|---|---|---|
| | | n | Median change | n | Median change | | |
| Step Velocity, m/s | | | | | | | |
| PSI [Lower Limbs] | From BL | 7 | -0.177 | 23 | -0.014 | 12.4 | 0.016 |
| PGI-S | From BL | 7 | -0.145 | 25 | -0.058 | 2.5 | 0.755 |
| PGI-C | From W72 | 77 | -0.021 | 48 | -0.004 | 5.0 | 0.176 |
| PGI-C | From BL | 82 | -0.079 | 53 | -0.049 | 1.6 | 0.174 |
| Step Velocity Scaled to Walking Time, m/s | | | | | | | |
| PSI [Lower Limbs] | From BL | 7 | -0.177 | 23 | -0.014 | 12.4 | 0.016 |
| PGI-S | From BL | 7 | -0.145 | 25 | -0.061 | 2.4 | 0.721 |
| PGI-C | From W72 | 77 | -0.023 | 48 | -0.007 | 3.4 | 0.236 |
| PGI-C | From BL | 82 | -0.109 | 53 | -0.049 | 2.2 | 0.138 |
| Step Duration, s | | | | | | | |
| PSI [Lower Limbs] | From BL | 7 | 0.06 | 23 | 0.02 | 3.0 | 0.229 |
| PGI-S | From BL | 7 | -0.02 | 25 | 0.04 | -0.5 | 0.273 |
| PGI-C | From W72 | 77 | 0.01 | 48 | 0.00 | NA | 0.471 |
| PGI-C | From BL | 82 | 0.02 | 53 | 0.02 | 1.0 | 0.376 |
| Step Length, m | | | | | | | |
| PSI [Lower Limbs] | From BL | 7 | -0.062 | 23 | 0.000 | -191.8 | 0.061 |
| PGI-S | From BL | 7 | -0.062 | 25 | -0.010 | 6.5 | 0.370 |
| PGI-C | From W72 | 77 | -0.011 | 48 | -0.010 | 1.1 | 0.821 |
| PGI-C | From BL | 82 | -0.040 | 53 | -0.018 | 2.2 | 0.176 |
| Total Number of Steps | | | | | | | |
| PSI [Lower Limbs] | From BL | 7 | -32.0 | 23 | -2.0 | 16.0 | 0.035 |
| PGI-S | From BL | 7 | -19.0 | 25 | -19.5 | 1.0 | 0.927 |
| PGI-C | From W72 | 77 | -3.0 | 48 | -1.0 | 3.0 | 0.282 |
| PGI-C | From BL | 82 | -13.0 | 53 | -13.5 | 1.0 | 0.377 |
| Total Distance Walked, m | | | | | | | |
| PSI [Lower Limbs] | From BL | 7 | -33.6 | 23 | -2.9 | 11.8 | 0.008 |
| PGI-S | From BL | 7 | -24.6 | 25 | -16.4 | 1.5 | 0.447 |
| PGI-C | From W72 | 77 | -1.8 | 48 | -1.1 | 1.6 | 0.498 |
| PGI-C | From BL | 82 | -10.5 | 53 | -8.2 | 1.3 | 0.271 |

From BL, from baseline; NA, not applicable; PGI-C, Patient Global Impression of Change; PGI-S, Patient Global Impression of Severity; PSI [Lower Limbs], Patient Global Impression of Severity for Lower Limbs; W72, Week 72.

**Supplementary Table S4. Qualitative study: baseline demographics and disease characteristics.**

| Variable | Concept elicitation (N = 40) | Cognitive debriefing (N = 16) |
|---|---|---|
| Female, n (%) | 25 (62.5) | 11 (68.7) |
| Age, years, mean (range) | 52.4 (28–68) | 49.4 (29 – 62) |
| MS type, n (%) | | |
| RRMS | 12 (30.0) | 9 (56.3) |
| SPMS | 8 (20.0) | 7 (43.8) |
| PPMS | 20 (50.0) | 0 (0) |
| Patient-reported EDSS | | |
| Mean (SD) | 5.0 (1.9) | 4.7 (2.2) |
| 0 – 2.5, n (%) | 7 (17.5) | 4 (25.0) |
| 3.0 – 5.5, n (%) | 15 (37.5) | 5 (31.3) |
| ≥ 6.0, n (%) | 18 (45.0) | 7 (43.7) |

EDSS, Expanded Disability Status Scale; PPMS, primary progressive MS; RRMS, relapsing-remitting multiple sclerosis; SPMS, secondary progressive multiple sclerosis.

**Supplementary Table S5. Concept elicitation: example patient statements on how MS affects their walking ability.**

| Gait concept impacted by the symptoms described by the patient | Digital gait measure that captures the symptoms described by the patient | Symptoms described by the patient |
|---|---|---|
| Ability to walk fast | • Step Velocity<br>• Step Velocity Scaled to Walking Time<br>• Step Duration | • "I have to walk a lot slower, and I seem to sometimes have to almost limp a little bit."<br>• "I used to walk 10-minute miles, and now it's extremely slow."<br>• "I try not to speed up. Because for some reason, I feel that, if I walk fast, I drag, I would say, the tip of my foot. And it scrapes the bottom."<br>• "For some reason, when I'm trying to walk fast, it drags."<br>• "I use a cane, and I walk slow."<br>• "Other than you know, using the cane, you know. Because it makes…it slows me down."<br>• "The fact that I can't walk normal. I can't walk with any quickness or anything, any speed."<br>• "Probably, because that's why I go slower. You know, I try to go real fast and I can't do that fast so maybe there is a little muscle weakness."<br>• "For some reason, when I'm trying to walk fast, it drags. And that's what causes me to trip. And like I said, I'm afraid of actually falling."<br>• "I walk, I exercise probably 5 or 6 days a week. I used to walk 10-minute miles, and now it's extremely slow."<br>• "[...] why am I dragging my foot? So, at the same time that I'm thinking about it, I try to correct myself. And that's when I notice that I have to take my time and walk." |
| Need to take rests while walking | • Step Velocity Scaled to Walking Time | • "Sometimes it even gets worse, and I have cramping, like really serious difficulty walking, and I just have to take a break for a little bit and sit down."<br>• "[...] after sitting for a few minutes, I often can't really walk that far, straight. I have to take breaks after a few moments."<br>• "Luckily, I live in a condo, everything's on one floor. But it's a big condo. So, walking from one end to the other, I often become winded or I have to stop, or I trip over something."<br>• "I couldn't walk as far as I used to be able to walk. I had to keep sitting down."<br>• "I try not to go for long distances of walks. I try to go short distance, and take my time. And if I see a place where I can sit, after I start…I, myself, start feeling a little bit tired, I try to find a place."<br>• "We always do it in 50 yard [inaudible] go around the track and stuff, and I take breaks. And I could walk a little bit and then sit down and take a break and walk some more." |
| | | • “I know that when I'm walking for long periods of time, I can't walk as much as most people without being tired or anything like that. So I have to sit down, just rest, even if it's for a few seconds.” |
| Step quality | • Step length | • “It’s just that sometimes I have trouble moving my legs, it’s just as simple as that, sometimes it’s a short taking steps, and then the fatigue, I guess you can’t have fatigue in your legs, but it’s basically the stiffness that bothers me the most” |
| Distance, stamina, and endurance | • Total Number of Steps<br>• Total Distance Walked | • “Now I used to take like long walks. Now I can’t. I, you know, for long walks, yeah.”<br>• “I can't walk as far as I used to.”<br>• “Then it got to where I could only walk a short distance had…the wheelchair had to be in the same room with the bed, and then it got to where I can't even transfer.”<br>• “I usually walk 3 miles a day. I couldn’t do that today. I did that even when I was first diagnosed. It was probably 4 or 5 years before I stopped walking.”<br>• “There’s some days where I can go maybe like 500, maybe 600, 700 feet or something.”<br>• “I could walk for a longer distance before having to stop. Now I can’t walk that far, like walking from one end of my place to the other.”<br>• “I’ve noticed that maybe one month I was able to walk 200 feet without losing balance or if like I said I’m looking at the ground or I’m familiar with the ground. And now even if I’m familiar with the terrain, it seems to be becoming less and less of a distance that I can walk.”<br>• “I can't walk more than about 30 or 40 feet and, at that, with a cane.”<br>• “[...] my car is parked about 40 feet from my apartment. That's a walk that sometimes I don't even want to try to make.”<br>• “[...] in any given day I may walk less than 100 feet.”<br>• “I can't walk. My grandson years ago told me, grandpa I wish you could run, and I said I can run, just can't do it for a long distance.”<br>• “Yeah, if I stand too long, or if I walk longer than to the end of my driveway, I guess, it’s…then it would be an issue, because then I start getting fatigued in my legs [...].”<br>• “I used to walk every day down the cul-de-sac and back, and I used to walk a mile and a half, 2 miles, every day, is what it totalled up to. And now I don’t even do that, in the morning.”<br>• “Oh, now. I’d say the farthest, maybe 1/4 mile or an 1/8 of a mile.” |

**Supplementary Table S6. Cognitive debrief: how daily activities are affected by lower limb symptoms that impact the performance on the 2MWT.**

| Daily activity | Impact of 2MWT-associated symptoms on daily activities as reported by patients |
|---|---|
| Walking | • "My most heart-breaking symptom is not being able to walk… Walked home for lunch, walked back to campus… I didn't think anything of when I was walking everywhere."<br>• "Walking in the store with my husband or walking into church or walking around the house." |
| Completing everyday routine | • "Walking, going places and doing things. It pretty much encompasses everything. Getting up going to the bathroom, everything."<br>• "I mean just vacuuming requires what you just had me do." |

2MWT, Two-Minute Walk Test.

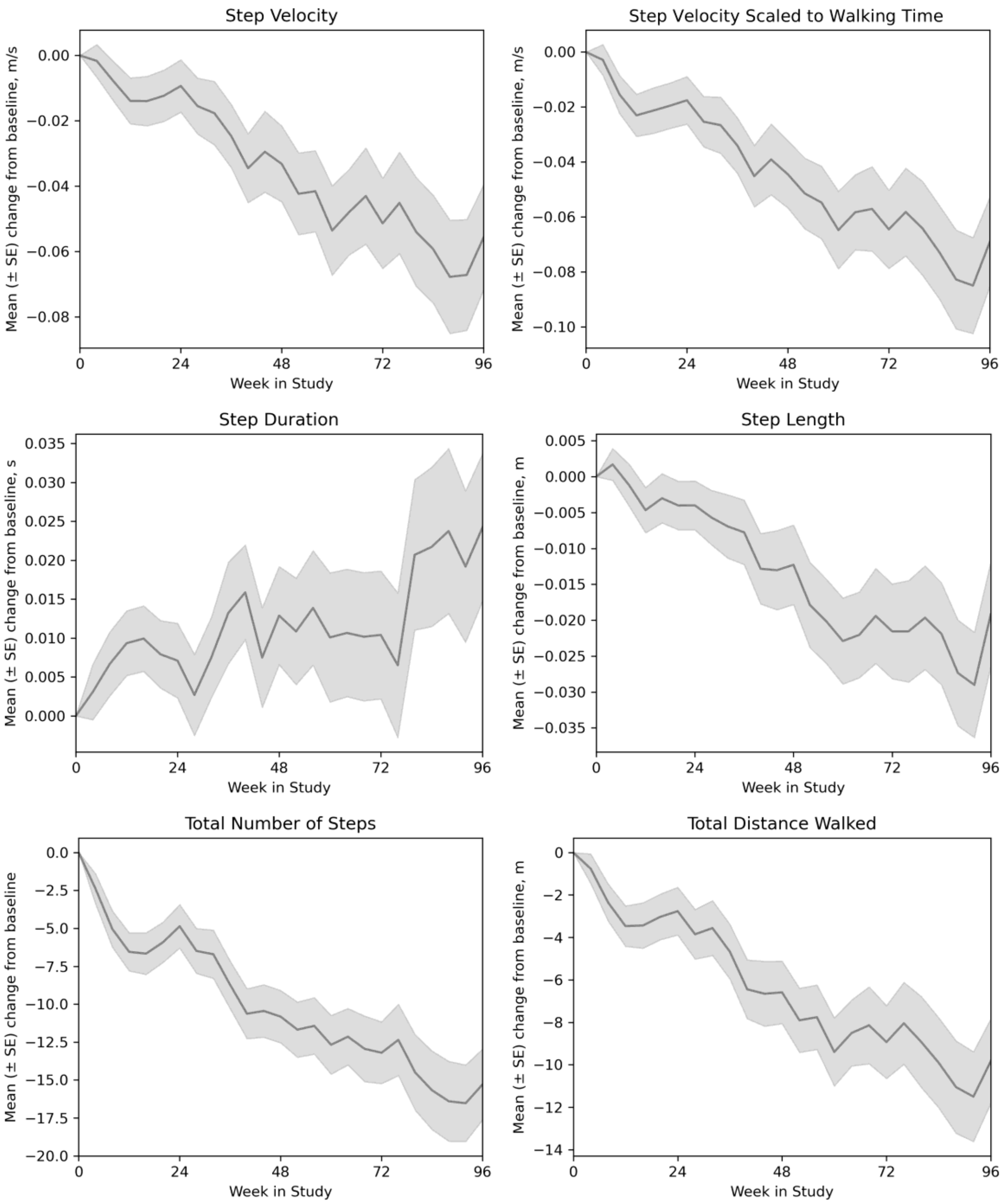

**Supplementary Figure S1.** Trajectories of digital gait measures from baseline up to Week 96.

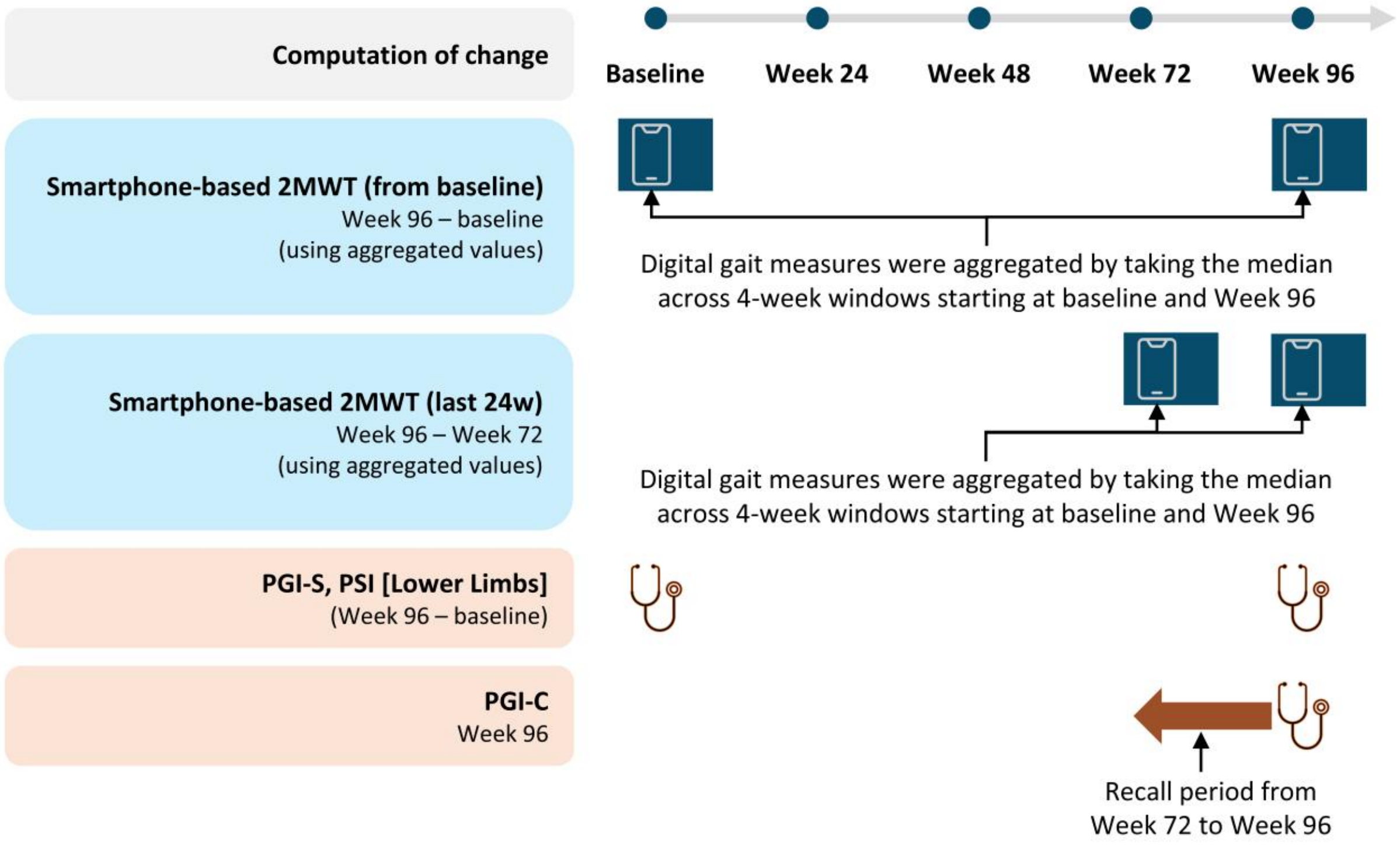

**Supplementary Figure S2.** The change from baseline at Week 96 was computed either as a change between aggregated values (smartphone-based 2MWT), a simple change (PGI-S and PSI [Lower Limbs]), or by taking the absolute score obtained at Week 96 (PGI-C). To accommodate the recall period on the PGI-C, changes on digital gait measures were computed both from baseline and from Week 72 (“last 24w”).
From BL, from baseline; last 24, last 24 weeks (from Week 72); PGI-C, Patient Global Impression of Change; PGI-S, Patient Global Impression of Severity; PSI [Lower Limbs], Patient Global Impression of Severity for Lower Limbs.

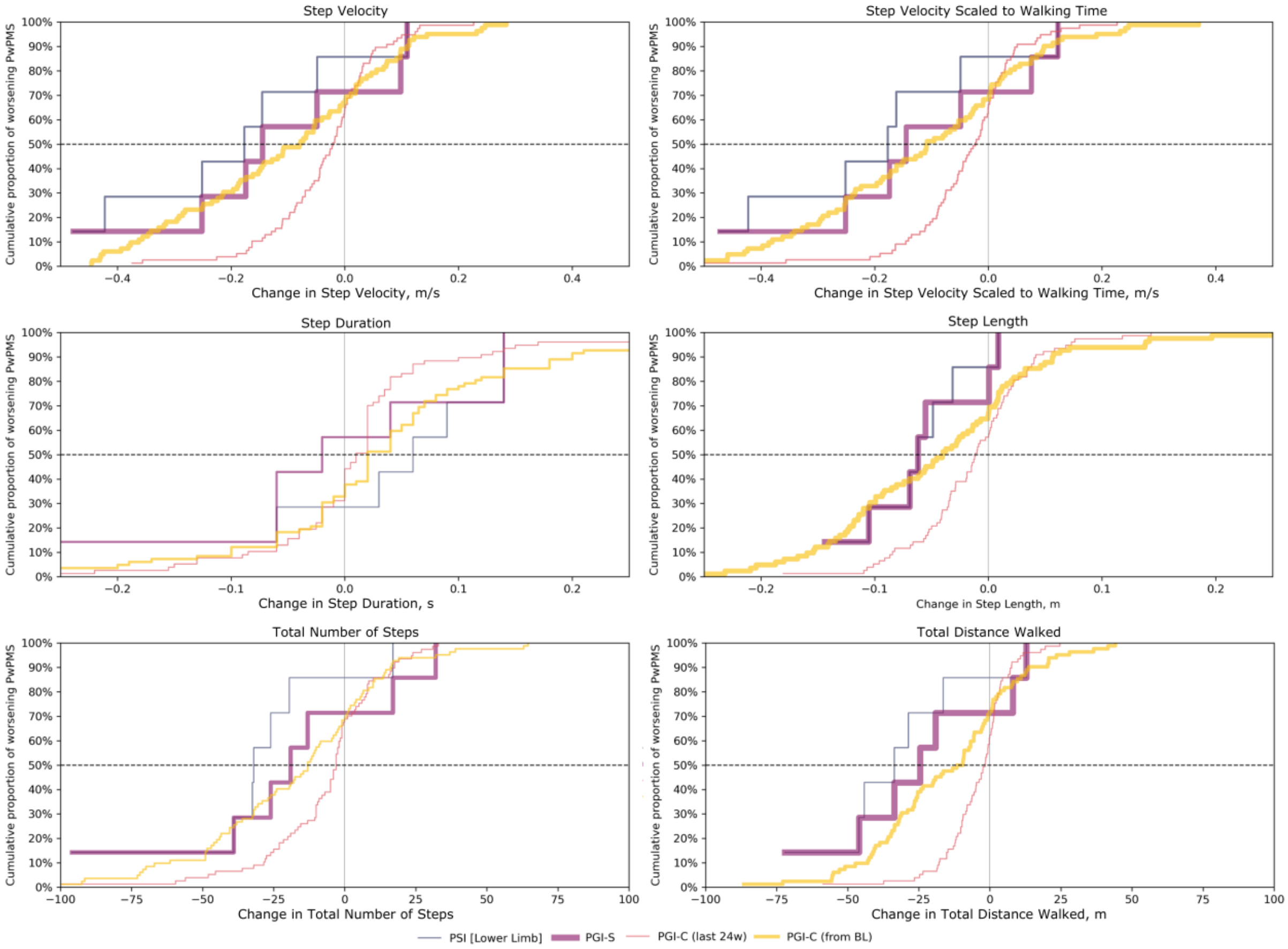

**Supplementary Figure S3:** Empirical cumulative distribution functions (ECDFs) for changes on digital gait measures from baseline at Week 96 are shown for worsening groups defined by thresholds for supplementary anchors (PSI, PGI-S, and PGI-C scales). Each curve represents the subset of patients who met the respective anchor-defined threshold for worsening reported in Table 1. The line width of each ECDF corresponds to the strength of the Spearman's rho longitudinal correlation between the digital measure and the respective anchor reported in Supplementary Table S2.